\documentclass[aps,showpacs, prl,twocolumn,superscriptaddress]{revtex4}%
\usepackage{epsfig,dsfont,amssymb,amsmath,amsthm,amsfonts,amsbsy,mathrsfs}
\usepackage{graphicx}
\usepackage{amsmath}
\usepackage{amssymb}
\begin{document}
\title{Pseudo-crystals of the group 14 elements with both 5-fold
central rotation symmetry and divisional translation symmetry}
\author{Chaoyu He}
\affiliation{Hunan Key Laboratory for Micro-Nano Energy Materials
and Devices, Xiangtan University, Hunan 411105, P. R. China;}
\affiliation{Laboratory for Quantum Engineering and Micro-Nano
Energy Technology, Xiangtan University, Xiangtan 411105, China.}
\author{Jianxin Zhong}
\email{jxzhong@xtu.edu.cn}\affiliation{Hunan Key Laboratory for
Micro-Nano Energy Materials and Devices, Xiangtan University, Hunan
411105, P. R. China;} \affiliation{Laboratory for Quantum
Engineering and Micro-Nano Energy Technology, Xiangtan University,
Xiangtan 411105, China.}
\date{\today}
\begin{abstract}
Crystals are the materials which can be described by uniform
periodic lattices. Traditionally, only the 1-, 2-, 3-, 4- and 6-fold
rotation symmetries are allowed in crystals because other n-fold
rotation symmetries are forbidden by the fully periodic translation
symmetry. Materials containing rotation symmetries forbidden in
crystals can be optionally described as quasicrystals with
aperiodicity. The theoretical predictions and experimental
discoveries of quasicrystals have enriched the knowledge of
crystallography and topology, and expanded the capability of growing
novel materials. Mathematically, one can fill the whole space with
quasi-uniform lattices through divisional translations and arbitrary
n-fold central rotations, including the 5-fold symmetry. However, it
is unknown if such pseudo-crystals with 5-fold symmetry exist in
reality. Here we propose a generalized crystal-prediction-method
which can be used to search for potential pseudo-crystals possessing
both arbitrary n-fold rotation symmetry and divisional translation
symmetry. Successful examples with 5-fold central rotation symmetry
based on the group 14 elements are discussed. We expect timely
experimental explorations on such peculiar structures, especially on
the very promising clathrate phases of the group 14 elements with
only slightly larger binding energies compared with the diamond phases.\\
\end{abstract}
\maketitle
%section{Introduction}
\indent Crystals are a class of materials which can be described by
periodic lattice with translation symmetry. According to the
well-know theorems of crystallography derived nearly two centuries
ago, only the 1-, 2-, 3-, 4- and 6-fold symmetries are allowed in
crystals because other n-fold symmetries, such as 5- and 7-fold
ones, are forbidden by the fully periodic translation symmetry. The
absence of 5-fold symmetry in periodic lattices had been mysterious
in history for scientists. Kepler explored some special arrangements
of planar pentagons \cite{1} which can be viewed as preforms of
Penrose's aperiodic tilings \cite{2}. Their endeavors provide
important mathematical theory to describe a new type of materials
defined as quasicrystals \cite{3} with aperiodicity and the rotation
symmetries forbidden in crystals. Since 1984 when Shechtman et al.
first discovered \cite{4} the crystal-like diffraction pattern with
forbidden icosahedral symmetry from aluminum-manganese alloys, the
mathematical topologies of pentagons derived from Kepler \cite{1}
and Penrose \cite{2} have became more and more popular \cite{5} in
mathematics, crystallography and physics, and the
quasi-crystallography has been developed into an elaborate
disciplines \cite{3}. The theoretical predictions and experimental
discoveries of quasicrystals have significantly enriched our
knowledge of crystallography and topology, and the Nobel Prize in
Chemistry 2011 was awarded to Shechtman for the discovery of
quasicrystals. Today, it is clear that one can only have periodic
crystals and aperiodic quasicrystals, and any crystalline phases
cannot possess both fully translation symmetry and 5-fold rotation
symmetry. Nevertheless, we note that one can mathematically fill the
whole space with quasi-uniform lattices through divisional
translation operations and arbitrary n-fold central rotation
operations. For example, Fibonacci pentilings containing 5-fold
symmetry can be described by quasi-uniform parallelogram lattices
through divisional translation operations and 5-fold rotation
operations \cite{5}. Here, we focus on the important question that
"Is there any physically realistic materials containing both
divisional translation symmetry and 5-fold rotation symmetry can be
theoretically constructed and experimentally touchable?". We expect
such a novel topological manner of materials as new optional
understandings \cite{p1,p2} of the potential experimental
discoveries of 5-, 8-, 10- and other higher n-fold
diffraction-patterns. We propose a generalized
crystal-prediction-method for the purpose of searching for potential
crystalline networks containing both divisional translation symmetry
and arbitrary n-fold rotation symmetry. With this method, we
successfully realized some promising pseudo-crystals possessing
both 5-fold central rotation symmetry and divisional translation symmetry,
which can be potentially formed by the group 14 elements.\\
%*************************figure 1****,bb=0 0 1750 400*********************************************************************
\begin{figure}
\center
\includegraphics[width=3.5in]{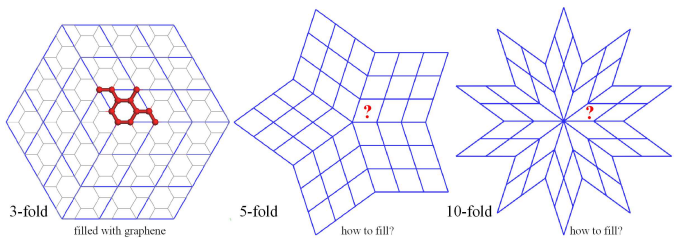}\\
\caption{Examples of filling the whole space with quasi-uniform
lattice cells through divisional translation-operation and 3-, 5-,
10-fold rotation-operations. Lattice cells are quasi-uniform because
that their local environments are different.}\label{fig1}
\end{figure}
\indent In traditional crystal prediction methods
\cite{M1,M2,M3,M4}, one constructs the initial testing phase of
elements or compounds through randomly filling specially defined
lattice cell with corresponding symmetry operations from the space
group and then fill the whole space with the constructed unit cell
through only the translation operations to form a crystal. In fact,
under some special conditions we can fill the whole space with the
constructed unit cell through the combination of translation
operations and n-fold rotation operation. As shown in Fig.1 for the
two-dimensional case, it is mathematically acceptable when the
lattice cell is non-triclinic and the third lattice angle is
2$\pi$/n. Taking the 5-fold (10-fold) one as example, we can fill
the 1/5 area of the whole space with the 72$^o$ (36$^o$) monoclinic
cells through translation operation and the rest 4/5 areas with
C$^1_5$, C$^2_5$, C$^3_5$ and C$^4_5$ rotation operations. That is
to say, we can mathematically construct crystals with both
translation symmetry and arbitrary n-fold central rotation symmetry.
For the physically acceptable substantiality, we prefer to use the
perfect and simple lattice of graphene as an example for the 3-fold
condition. As indicated in Fig.1, we can represent graphene with the
120$^o$ hexagonal lattice cell (with six carbon atoms) through
translation operations and 3-fold rotation operations. We can also
describe graphene with the same 120$^o$ hexagonal lattice
cell through only translation operations in a traditional way.\\
%*************************figure 3*************************************************************************
\begin{figure}
\includegraphics[width=3.5in]{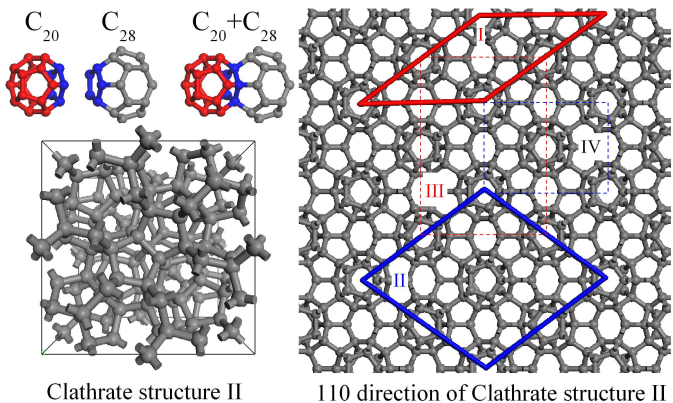}\\
\caption{Fundamental structural units of cage 20 and cage 28, cubic
crystalline cell and the 110 direction view of clathrate structure
II. Four optional supercells for representing clathrate structure II
are indicated as I, II, III and IV.}\label{fig3}
\end{figure}
%============================
\indent To construct a physically acceptable pseudo-crystal with
both divisional translation symmetry and arbitrary n-fold central
rotation symmetry, we need a proper arrangement of atoms or
molecules in the unit cell, which ensures that neither translation
operation nor rotation operation creates unreasonable configuration,
especially on the areas of cell boundaries. Such a special
physically acceptable arrangements of atoms or molecules can be
easily identified by well-developed computer programs for topology
analysis (such as TOPOS, Topological Databases and Topological Types
Observed, http://www.topos.ssu.samara.ru) and physically optimized
by first-principles calculations through a special
cell-shape-keeping procedure. In summary, we can predict pseudo
crystalline phases of elements or compounds with both divisional
translation symmetry and arbitrary n-fold central rotation symmetry
through: 1) randomly fill the special unit cell with atoms or
molecules, 2) structurally filter reasonable arrangements of atoms
or molecules, 3) physically optimize the unit cell and its inner
atomic positions by first-principles calculations, and 4) extend the
unit cell to crystalline form through divisional translation
operation and corresponding n-fold rotation operation. With this
method, we are able to predict many new crystalline phases of
elements or compounds with both divisional translation symmetry and
arbitrary
n-fold central rotation symmetry.\\
%*************************figure 2*************************************************************************
\begin{figure}
\includegraphics[width=3.5in]{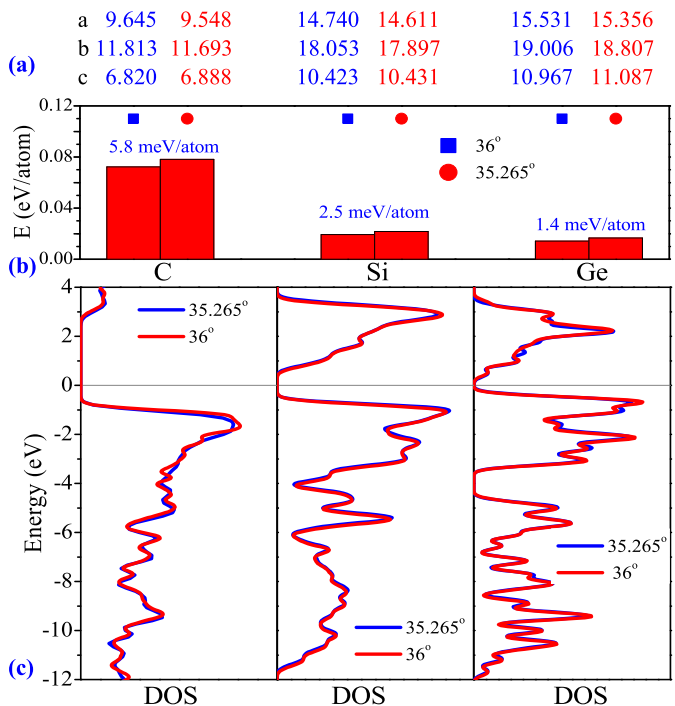}\\
\caption{Comparing in lattice constants of lattice-I and
36$^o$-lattice-I (a), average-energies and deformation energies (b)
of lattice-I and 36$^o$-lattice-I, and the calculated density of
states (c) of lattice-I and 36$^o$-lattice-I for C, Si and
Ge.}\label{fig2}
\end{figure}
%++++++++++++++++++++
\indent Except for the two-dimensional graphene for the 3-fold
condition discussed above, we find that the clathrate structure of
type II can be considered as an excellent example for filling the
three-dimensional space with divisional translation operations and
5-fold central rotation operations. Fig.2 shows the fundamental
structural units, cage 20 and cage 28, of the clathrate structure
II. Clathrate structure II can be realized in crystalline forms of
elements Si and Ge with or without guest atoms in the cage positions
\cite{c1,c2,c3,c4,c5}. It has also been discovered in crystalline
forms of Tin \cite{c6} and hydrates \cite{c7,c8}, in which guest
atoms are popular. Clathrate structure II containing structural
units of cage 20 and cage 28 can be described by a face-center cubic
lattice with 136 vertexes (Fcc136), in which only three vertexes are
not equivalent due to the high symmetry of its space group of Fd-3m
(227). In fact, clathrate structure II can also be optionally
described by other forms of lattices, such as the monoclinic I, II
and the rectangular III and IV as indicated in Fig.2. The third
angles of the monoclinic lattice-I and lattice-II are 35.265$^o$ and
75.530$^o$, respectively, very close to 36$^o$ and 72$^o$
(Lattice-II is a $\surd$2$\times$$\surd$2 supercell of lattice-I).
Interestingly, we find that applying the 5-fold rotation operation
(10-fold rotation reversion) on lattice-II (lattice-I) creates no
any unreasonable interfacial configurations if we impose its third
lattice angle to be 72$^o$ (36$^o$). This suggests that we can
theoretically construct a promising crystalline network containing
both divisional translation symmetry and 5-fold (10-fold) central
rotation symmetry based on 72$^o$-lattice-II (36$^o$-lattice-I),
which can be potentially achieved by the group 14 elements Si and Ge
\cite{c1,c2,c3,c4,c5}. Element Tin \cite{c6} and compound hydrates
\cite{c7,c8} are also proper candidates but we focus on the group
14 elements C, Si and Ge in this work as examples.\\
%++++++++++++++++++++
\begin{figure}
\includegraphics[width=3.5in]{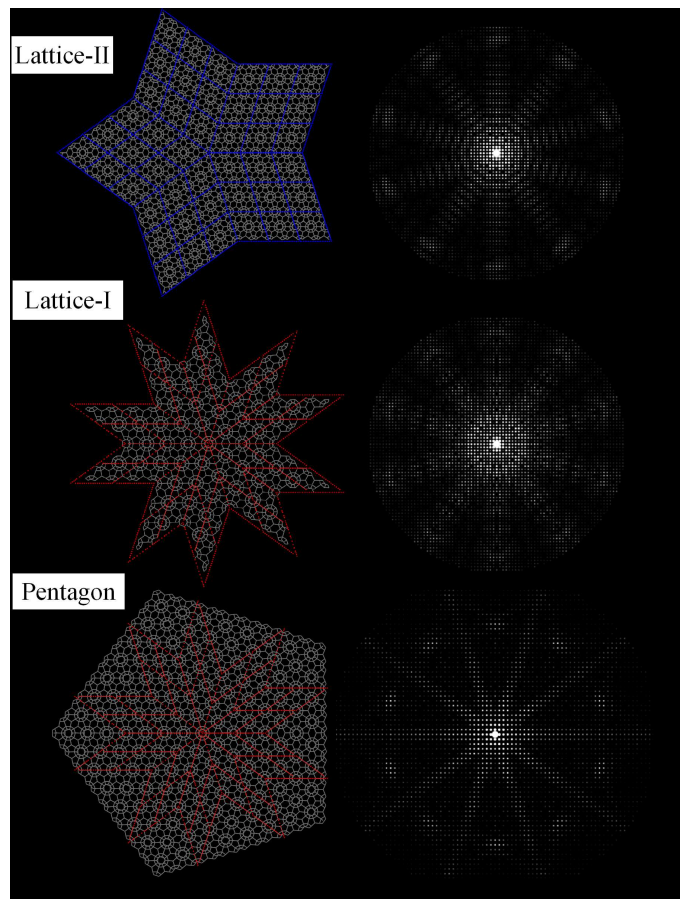}\\
\caption{Simulated electron-diffraction-patterns based on element Ge
with finite size in shapes of 5-pointed star, 10-pointed star and
pentagon, respectively.}\label{fig4}
\end{figure}
%============================================================
\indent We then performed cell-shape-keeping optimizations of the
36$^o$-lattice-I structures in the situations of elements C, Si and
Ge, respectively and then constructed the corresponding crystals of
72$^o$-lattice-II through extending the optimized ones of
36$^o$-lattice-I to $\surd$2$\times$$\surd$2 supercells. The
optimized lattice constants for crystals of 36$^o$-lattice-I and
lattice-I (35.265$^o$) for C, Si and Ge are summarized in Fig.3.
Comparing the lattice constants, we can see that the shear
deformation from 35.265$^o$ to 36$^o$ slightly decreases the lattice
constants of a and b but increases the lattice constant c. The small
deformations in lattice-I for elements C, Si and Ge require very
small deformation energies, indicating that the structures of
36$^o$-lattice-I are physically realistic and experimentally
touchable. In Fig.3, we also summarize the average energies of
lattice-I and 36$^o$-lattice-I for C, Si and Ge, respective to their
corresponding ground states in the diamond-lattice. The
corresponding deformation energies from lattice-I to
36$^o$-lattice-I for C, Si and Ge are also summarized. For carbon,
we can see that lattice-I possesses relatively high energy ($\sim$75
meV/atom) respective to diamond-lattice. This may be the reason that
carbon-version clathrate structure II has not been experimentally
reported. The deformation energy of lattice-I to 36$^o$-lattice-I of
5.8 meV/atom is very small to be realized by temperature or other
growth conditions. The average energy of Si-version (Ge-version)
lattice-I respective to its corresponding ground state of
diamond-lattice is 19 meV/atom (14 meV/atom). Such a small energy
difference indicates a high probability of co-existences of Si (Ge)
in lattice-I and diamond-lattice, which is in good agreement with
the fact that Si and Ge clathrate II structures have been
successfully discovered in experiments \cite{c1,c2,c3,c4,c5}.
Moreover, the deformation energy of Si (Ge) lattice-I from
35.265$^o$ to 36$^o$ is only 2.5 meV/atom (1.4 meV/atom), indicating
that Si (Ge) phases in perfect lattice-I and 36$^o$-lattice-I are
nearly energy degenerated. From the calculated density of state of
these phases as shown in Fig.3, we can see that the slight
deformations in these crystalline phases of C, Si and Ge slightly
affect their electronic properties. The density of states of
crystals of lattice-I and crystals of 36$^o$-lattice-I are nearly
degenerate too. Experimentally, Si and Ge phases in the perfect
lattice-I (equal to clathrate structure II) have been successfully
discovered \cite{c1,c2,c3,c4,c5}. We expect their corresponding
36$^o$-lattice-I can be discovered in future experiments, especially
in the special topological manner with 5-fold central rotation
symmetry as discussed below.\\
%=============================
\indent We then construct physically realistic and experimentally
touchable pseudo crystalline phases of C, Si and Ge with 5-fold
central rotation (10-fold rotation-reversion) symmetry based on the
optimized 72$^o$-lattice-I (36$^o$-lattice-II). As shown in Fig.4,
we extend the 72$^o$-lattice-I (36$^o$-lattice-II) through
divisional translation symmetry operations and 5-fold rotation
symmetry operations (10-fold rotation reversion) to form a 5-pointed
(10-pointed) star with finite size. Although 36$^o$-lattice-I and
72$^o$-lattice-II form different shapes and edge shapes, they are
topologically same pseudo-crystals when they topologically fill
fully the infinite three-dimensional space. For example, both can be
cut to a uniform shape of pentagon as shown in Fig.4. The
electron-diffraction-patterns of Ge in the shapes of 5-pointed star,
10-pointed star and pentagon were simulated and are shown in Fig.4.
The corresponding diffraction-patterns for C and Si are very similar
to those for Ge (not shown in this letter). From Fig.4, we can see
that the diffraction-patterns based on the 5-pointed star,
10-pointed star and pentagon are some different to each other in
view of their finite sizes and different edge shapes. However, all
of them contain clear 5-fold symmetry and 10-fold symmetry. These
diffraction-patterns can be used to identify this novel type of
crystals of C, Si and Ge in future experiments. We expect
timely experimental explorations on such peculiar structures.\\
%***********************************
\indent In summary, we have proposed a generalized
crystal-predication-method to predict potential pseudo-crystals
possessing both arbitrary n-fold central rotation symmetry and
divisional translation symmetry through the combination of the
traditional space-groups and the point-groups of regular polygons.
With this method, we can predict many pseudo crystalline phases of
elements or compounds with both divisional translation symmetry and
arbitrary n-fold central rotation symmetry, which are significant
for extending our the knowledge of crystallography and topology and
can be considered as new options for understanding potential
discoveries of 5-, 7- and higher n-fold diffraction patterns in
future experiments. Based on the experimentally touchable Si and Ge
phases in clathrate structure II, pseudo crystalline forms of C, Si
and Ge with 5-fold and 10-fold symmetry are predicted.\\
%\section*{Acknowledgement}
\indent This work is supported by the National Natural Science
Foundation of China (Grant Nos. 11074211 and 51172191), the National
Basic Research Program of China (2012CB921303), and the Hunan
Provincial Innovation Foundation for Postgraduate (Grant No. CX2013A010).\\
%+++++++++++++++++++++++++++++++++++++++++++++++++++++++++


\begin{thebibliography}{29}
\bibitem{1} Senechal, M. (1995) Quasicrystals and Geometry (Cambridge Univ. Press, Cambridge, U.K.)
\bibitem{2} Penrose, R. (1979) Math. Intelligecer 2, 32-37.
\bibitem{3} Levine, D. and Steinhardt, P. J. (1984) Phys. Rev. Lett. 53, 2477-2480
\bibitem{4} Shechtman, D., Blech, I., Gratias, D. and Cahn, J. W. (1984), Phys. Rev. Lett. 53,1951-1953.
\bibitem{5} Caspar, D. L. D. and Fontano, E. (1996), Proc. Natl. Acad. Sci. USA, 93, 14271-14278.
\bibitem{p1} Pauling, L (1985), Nature (London), 317, 512.
\bibitem{p2} Pauling, L (1987), Phys. Rev. Lett., 58, 365.
\bibitem{M1} Strong, S. T., Pickard, C. J. Milman, V. Thimm, G. and Winkler, B. (2004), Phys. Rev. B: condens. Matter Mater. Phys. 70, 045101-045107.
\bibitem{M2} Oganov, A. R. And Glass, C. W. (2006), J. Chem. Phys. 124, 244704-244715.
\bibitem{M3} Wang, Y. C. Lv, J. Zhu, L. and Ma. Y. M. (2010), Phys. Rev. B: condens. Matter Mater. Phys. 82, 094116-094122.
\bibitem{M4} Pickard, C. J. and Needs, R. J. (2011), J. Phys.: Condens. Matter, 23 053201.
\bibitem{c1} Gryko, J. Mcmillan, P. F. Marzke, R. F. Ramachandran, G. K. Patton, D. Deb, S. K. And Sankey, O. F., (2000), Phys. Rev. B, 62, R7707.
\bibitem{c2} Nolas, G. S. Kendziora, C. A. Gryko, J. Dong, J. J. Myles, C. W. Poddar, A. And Sankey, O. F. (2002), J. Appl. Phys. 92, 7225.
\bibitem{c3} Guloy, A. M. Ramlau, R., Tang, Z. J. Schnelle, W. Baitinger, M. and Grin, Y, (2006), Nature, 443,320.
\bibitem{c4} Fassler, T. F., (2007), Angew. Chem. Int. Ed. 46, 2572.
\bibitem{c5} Beekman, M. And Nolas, G. S. (2006), Physica B, 383, 111-114.
\bibitem{c6} Schafer, M. C. and Bobev, S. (2013), J. Am. Chem. Soc, 135, 1696-1699.
\bibitem{c7}Ripmeester, J. A., Tse, J. S. Ratcliffe, C. I. and Powell, B. M. (1987), Nature, 325, 135.
\bibitem{c8} Mao, W. L. Koh, C. A. and Sloan, E. D. (2007), Phys. Today 10, 42.
\end{thebibliography}
\end{document}